# Revisiting Mars' Induced Magnetic Field and Clock Angle Departures under Real-Time Upstream Solar Wind Conditions

Zhihao Cheng[1], Chi Zhang[1,*], Chuanfei Dong[1,*], Hongyang Zhou[1], Jiawei Gao[1], Abigail Tadlock[1], Xinmin Li[1], and Liang Wang[1]

[1]Center for Space Physics and Department of Astronomy, Boston University, Boston, MA, USA

*Corresponding author: Chi Zhang (zc199508@bu.edu) and Chuanfei Dong (dcfy@bu.edu)

**Key Points:**

- We present a statistical study of Martian induced magnetic fields and clock angle departures under real-time solar wind conditions.
- The induced magnetic fields are controlled by solar wind dynamic pressure and the magnitude of IMF.
- Clock angle departures in the magnetosheath are generally small, but they vary depending on solar wind conditions.

**Abstract**

Mars lacks a global intrinsic dipole magnetic field, but its interaction with the solar wind generates a global induced magnetosphere. Until now, most studies have relied on single-spacecraft measurements, which could not simultaneously capture upstream solar wind conditions and the induced magnetic fields, thereby limiting our understanding of the system. Here, we statistically re-examine the properties of Mars' induced magnetic field by incorporating, for the first time, real-time upstream solar wind conditions from the coordinated MAVEN and Tianwen-1 observations. Our results are show that both solar wind dynamic pressure and the interplanetary magnetic field (IMF) magnitude enhance the strength of the induced magnetic field, but they exert opposite effects on the compression ratio: higher dynamic pressure strengthens compression, while stronger IMF weakens it. The induced field is stronger under quasi-perpendicular IMF conditions compared with quasi-parallel IMF, reflecting a stronger mass-loading effect. We further investigate the clock angle departures of the induced fields. They remain relatively small in the magnetosheath near the bow shock, increase gradually toward the induced magnetosphere, and become significantly larger within the induced magnetosphere. In addition, clock angle departures are strongly enhanced under quasi-parallel IMF conditions. Their dependence on upstream drivers further shows that, within the magnetosheath, clock angle departures are minimized under low dynamic pressure, high IMF magnitude, and low Alfvén Mach number conditions. These results may enhance our understanding of solar wind interaction with Mars, and highlight the critical role of multi-point observations.

**Plain Language Summary**

Understanding how the solar wind interacts with Mars provides important clues to the planet's evolution from a warm, wet environment to the cold, dry world we see today. Mars lacks a global intrinsic magnetic field, but the interaction between the solar wind and its upper atmosphere produces a global induced magnetosphere. The large-scale configuration of this induced magnetosphere is therefore strongly controlled by solar wind conditions. Until now, however, most studies have relied on single-spacecraft measurements, which could not simultaneously capture both upstream solar wind conditions and the induced magnetic fields. In this study, we use the joint observations from NASA's MAVEN mission and China's Tianwen-1 orbiter, providing for the first time an opportunity to examine Mars' induced magnetosphere under real-time upstream solar wind conditions. Our results may advance understanding of how the solar wind shapes Mars' space environment and highlight the unique value of multi-spacecraft observations for studying solar wind–planet interactions.

**1 Introduction**

Despite the absence of a global intrinsic dipole field, the interaction between the Martian upper atmosphere and the external solar wind, together with the interplanetary magnetic field (IMF), gives rise to a global induced magnetosphere (e.g., Bertucci et al., 2011; Luhmann et al., 2004; Zhang et al., 2022, 2023, 2025a, 2025b; Halekas et al., 2021; Dubinin et al., 2019; Gao et al., 2024; Ramstad et al., 2020). On a global scale, this system can be understood as the conducting obstacle of Mars' ionosphere moving through the solar wind. This motion induces electric currents within the ionosphere, while simultaneously allowing ionospheric plasma to gain energy from the solar wind and mass-load it, thereby bending and compressing the IMF. Together, these processes generate the induced magnetosphere that includes the bow shock, magnetosheath, and magnetotail, resembling, in many respects, the intrinsic magnetospheres of magnetized planets such as Earth. Since atmospheric ion acceleration and escape are strongly controlled by the surrounding electromagnetic fields (e.g., Dubinin et al., 2011; C. Dong et al., 2014; Zhang et al., 2024), understanding the structure and dynamics of the induced magnetic fields is therefore of fundamental importance.

Over the past decade, both simulations and observations have significantly advanced our understanding of the global structure of Mars' induced magnetosphere (e.g., Brain et al., 2006; Fang et al., 2018; C. Dong et al., 2015a, 2015b, 2018a, 2018b; Ma et al., 2002, 2018, 2019; Ramstad et al., 2020; Gao et al., 2024; Zhang et al., 2022, 2025a, 2025b). It has been established that the global pattern of induced magnetosphere is strongly controlled by the orientation of IMF or the solar wind motional electric field ($\vec{E}_{SW}$). In the +E hemisphere, where $\vec{E}_{SW}$ points away from the planet, the magnetic field exhibits a relatively regular configuration that closely follows the classical draping pattern characteristic of an induced magnetosphere. Moreover, the $\vec{E}_{SW}$-driven acceleration of planetary ions results in enhanced magnetic field compression in this hemisphere, producing a higher average field strength (Luhmann et al., 1985). By contrast, in the –E hemisphere where $\vec{E}_{SW}$ points toward the planet, the magnetic field structure is considerably more irregular (e.g., Harada et al., 2015; Zhang et al., 2022). Field lines there are more tightly wrapped around the planet and can even form loop-like structures above the polar regions (Chai et al., 2019; Dubinin et al., 2019).

In addition to IMF orientation, other solar wind parameters also exert a strong influence on the induced magnetosphere. For example, enhanced solar wind dynamic pressure compresses the induced magnetosphere and significantly increases the magnetic field strength (e.g., Zhang et al., 2022). When the IMF is nearly aligned with the solar wind velocity, the mass-loading effect weakens and the induced magnetosphere may degenerate (e.g., Q. Zhang et al., 2024). However, most previous studies have been limited to single-point measurements without real-time solar wind monitoring, and therefore assumed that solar wind conditions remain steady over the timescale of a spacecraft's orbital period. As a result, these studies may contain uncertainties, and their conclusions cannot be fully validated.

Since 2021, with the arrival of China's Tianwen-1 mission (Wan et al., 2020), Mars has been monitored by two orbiters capable of simultaneously measuring magnetic fields and plasma: NASA's MAVEN (Jakosky et al., 2015) and Tianwen-1. This dual-spacecraft configuration allows one orbiter to serve as a real-time solar wind monitor, while the other investigates the induced magnetic field inside Mars' interaction region. Such a setup provides a unique opportunity for more accurate characterization of the induced magnetosphere. Leveraging this capability, we present a statistical study to revisit the Martian induced magnetic fields using joint MAVEN and Tianwen-1 observations.

Previous studies have also suggested that the magnetosheath magnetic field clock angle can serve as a proxy for the upstream IMF (e.g., Y. Dong et al., 2019; Hurley et al., 2018; Fang et al., 2018). However, without real-time IMF measurements, this proxy method has not been fully validated. Thus, we also examine clock angle departures and their dependence on solar wind conditions to assess the accuracy of the proxy method.

**2 Datasets and Methods**

We use magnetic field data obtained by the MAG instrument on MAVEN (Connerney et al., 2015). The vector magnetic field measurements are available at both 32 Hz and 1 s resolutions; to reduce noise and suppress high-frequency fluctuations, we employ the 1 s resolution dataset. In addition, we incorporate magnetic field measurements from the Tianwen-1 Mars Orbiter Magnetometer (MOMAG; Y. Wang et al., 2023; Zou et al., 2023; G. Wang et al., 2024).

To determine whether the spacecraft are located in the upstream solar wind or in the downstream magnetosheath, we identify bow shock crossings that are typically characterized by sharp increases in magnetic field strength, enhanced wave activity, and ion heating (Gruesbeck et al., 2018; Zhang et al., 2025c), using magnetic field and ion measurements from the Solar Wind Ion Analyzer (SWIA; Halekas et al., 2015) onboard MAVEN and from the MOMAG and Mars Ion and Neutral Particle Analyzer (MINPA; Kong et al., 2020) onboard Tianwen-1.

We employed two Cartesian coordinate systems in this study. The first is the Mars Solar Orbital (MSO) coordinate system, where the $\vec{X}_{MSO}$ is along the vector from Mars to the Sun, $\vec{Z}_{MSO}$ is perpendicular to the orbital plane, and $\vec{Y}_{MSO}$ completes the right-handed system, closely aligned with the opposite direction of the orbital velocity vector. To better resolve the magnetic and velocity signatures of the plasma cloud and given that the configuration of the induced magnetosphere is strongly influenced by the solar wind and IMF orientation, we also adopted the Mars Solar Electric (MSE) coordinate system. In this system, $\vec{X}_{MSE}$ points antiparallel to the upstream solar wind flow, and $\vec{Z}_{MSE}$ points along the direction of the convection electric field ($\vec{E}_{SW}$) in the upstream solar wind, and the $\vec{Y}_{MSE}$ axis completes the right-handed system (e.g., Zhang et al., 2023). Thus, the $Z_{MSE} > 0$ hemisphere corresponds to the +E hemisphere, while $Z_{MSE} < 0$ denotes the -E hemisphere. Throughout this paper, we assume that the upstream solar wind flow is purely along the $-\vec{X}_{MSO}$ direction, which implies that the $\vec{X}_{MSO}$ and $\vec{X}_{MSE}$ are equivalent.

In our study, the clock angle is defined as the angle between the projected magnetic fields and $\vec{Z}_{MSO}$ in the $Y_{MSO} - Z_{MSO}$ plane (see Figure 1, Liu et al., 2021; Zhang et al., 2025d). It is calculated as $\phi = \arctan(B_y/B_z)$. The $\phi$ increases clockwise from the $+\vec{Z}_{MSO}$ toward $+\vec{Y}_{MSO}$. Accordingly, $\phi = 90°$ (or 270°) indicates that the projected magnetic field points toward $+\vec{Y}_{MSO}$ (or $-\vec{Y}_{MSO}$), while $\phi = 0°$ (or 180°) indicates that the projected magnetic field points toward $+\vec{Z}_{MSO}$ (or $-\vec{Z}_{MSO}$). The clock angle departure, $\Delta\phi$, is defined by $\phi_d - \phi_{imf}$, where $\phi_d$ is the clock angle of the downstream induced magnetic field and $\phi_{imf}$ is that of the upstream IMF. In this case, a positive (negative) $\Delta\phi$ means that the downstream induced magnetic field is rotated clockwise (counterclockwise) relative to the upstream IMF direction. Its magnitude ($|\Delta\phi|$) reflects the extent of deviation between the upstream IMF and downstream magnetic fields.

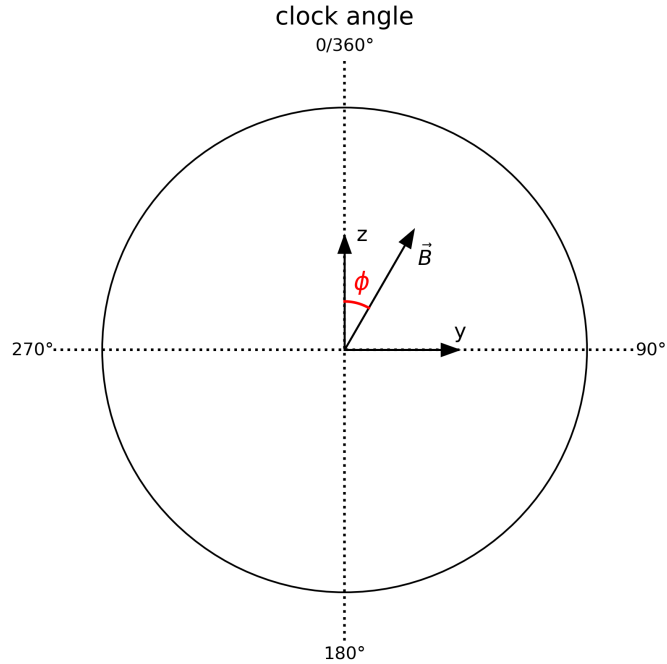

**Figure 1.** Definition of the magnetic field clock angle (Liu et al., 2021) in the MSO coordinate system. The clock angle is defined in the range of 0° to 360°, measured clockwise from the $+\vec{Z}_{MSO}$ direction.

## 3 Results

In this section, we first present a case study to illustrate how the downstream magnetic fields and $\Delta\phi$ are extracted and then provide statistical results to analyze their dependence on solar wind conditions.

### 3.1 Case Study

Figure 2 presents an example from 18:00 to 18:40 UTC on December 27, 2021. During this interval, Tianwen-1 was initially located in the magnetosheath, characterized by enhanced magnetic field strength and strong fluctuations (see Figure 2c). At approximately 18:04, Tianwen-1 crossed the bow shock and entered the solar wind region, where the magnetic field became weaker and more stable compared to the magnetosheath. Meanwhile, MAVEN was located in the induced magnetosphere before 18:20, as indicated by low ion fluxes and a relatively steady magnetic field (Figures 2d and 2e). After 18:20, MAVEN entered the magnetosheath, characterized by a marked increase in ion flux (100–2000 eV) and enhanced magnetic field turbulence.

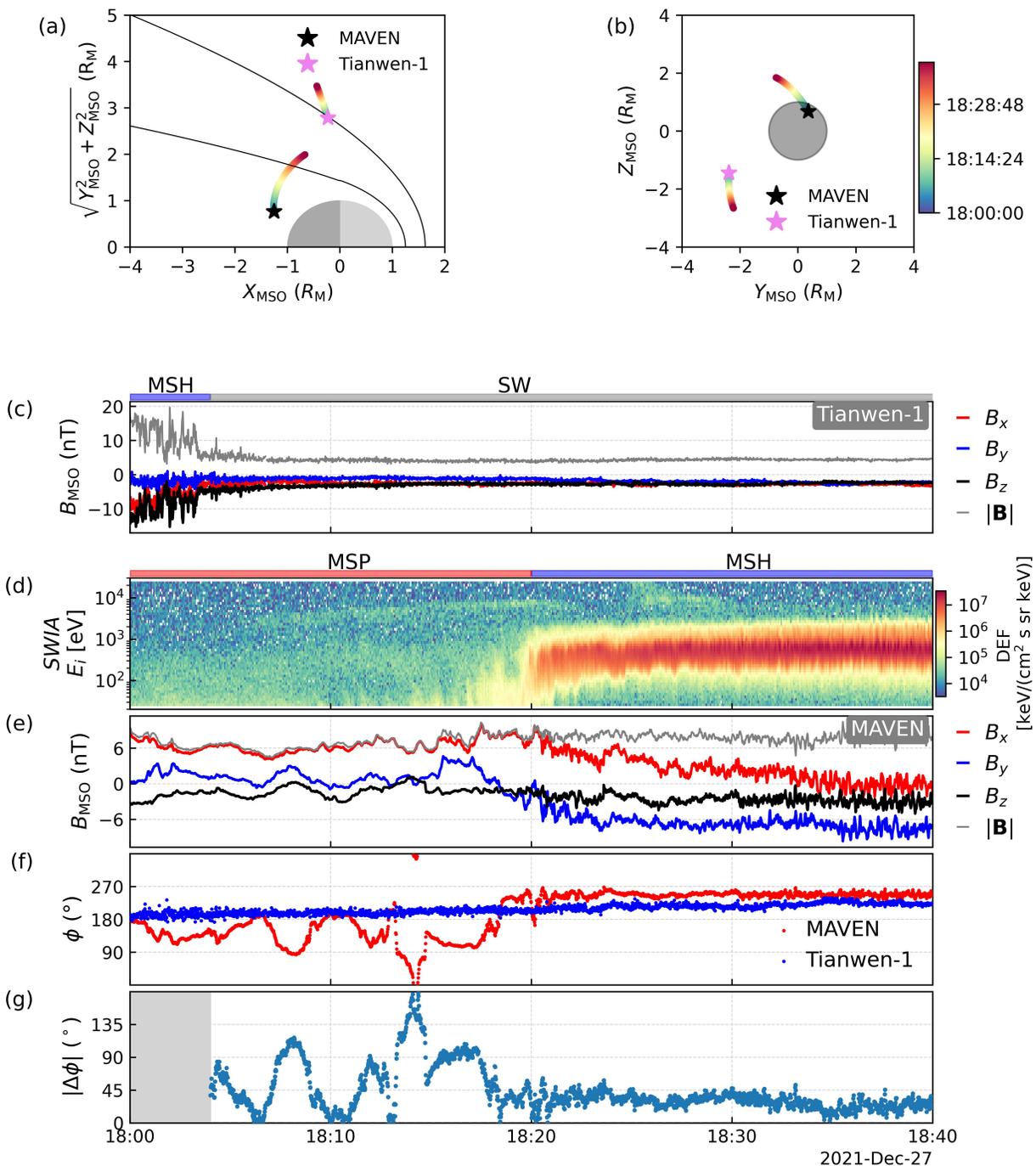

**Figure 2**. (a) Trajectories of MAVEN and Tianwen-1 in the $X_{MSO} - R_{MSO}$ plane, where $R = \sqrt{Y_{MSO}^2 + Z_{MSO}^2}$. The overlaid curves represent the nominal bow shock and the magnetic pile-up boundary (MPB) (Trotignon et al., 2006). (b) Trajectories of MAVEN and Tianwen-1 in the $Y_{MSO} - Z_{MSO}$ plane. (c) Magnetic field measured by Tianwen-1. (d) Ion energy spectrum observed by the SWIA instrument onboard MAVEN. (e) Magnetic field measured by MAVEN. (f) Clock angle of the magnetic field ($\phi$). (g) Absolute difference in clock angle ($|\Delta\phi|$) between Tianwen-1 and MAVEN. In the upper parts of panels (c) and (d), the blue shaded region labeled "MSH" denotes the magnetosheath, the red shaded region labeled "MSP" denotes the induced magnetosphere, and the gray shaded region labeled "SW" indicates the solar wind. In panel (g), the gray shaded interval marks the period when Tianwen-1 was not in the upstream region; $\Delta\phi$ during this interval are not considered.

Figure 2f compares the clock angles measured by MAVEN and Tianwen-1 during this interval. Before 18:20, when MAVEN was located in the induced magnetosphere, the clock angles show significant differences. In contrast, after 18:20, when MAVEN entered the magnetosheath, the clock angles observed by both spacecraft became much more consistent. This trend is further illustrated in Figure 2g, which shows the absolute clock angle departures, $|\Delta\phi|$. Since $\Delta\phi$ is defined as the difference between the downstream magnetic field (measured by MAVEN) and the upstream IMF (measured by Tianwen-1), we only consider $|\Delta\phi|$ after 18:04 when Tianwen-1 entered the upstream solar wind region. It is evident that $|\Delta\phi|$ is significantly larger and more variable when MAVEN is in the induced magnetosphere, while in the magnetosheath, $|\Delta\phi|$ remains below 45° and exhibits less fluctuation. These results suggest that the clock angle proxy method is unreliable within the induced magnetosphere, whereas it remains a reasonable approximation in the magnetosheath.

### 3.2 Statistical Results

To comprehensively evaluate the properties of induced magnetic fields and the clock angle departures, we conducted a statistical analysis using the MAVEN and Tianwen-1 datasets spanning from November 16, 2021, to August 31, 2022. The following criteria were applied to extract reliable data points:

(1) As previously discussed, we required that one spacecraft be located in the upstream solar wind region, while the other be located downstream (i.e., within the magnetosheath or induced magnetosphere).

(2) To minimize the influence of Martian crustal magnetic fields, which can significantly alter the magnetic fields, we exclude the data points where the observed magnetic field strength ($|B_{obs}|$) was less than ten times the strength ($|B_{model}|$) predicted by the crustal field model from Gao et al. (2021). Although this does not completely eliminate the influence of crustal fields, as they can still interact with the induced magnetic field, it serves as a practical first-order approximation to effectively suppress their impact in our analysis.

Based on the above criteria, we identified a total of 1.7 million data points. Figures 3a and 3b present the downstream magnetic field strength ($|\vec{B}_d|$) and the normalized field strength, expressed as $|\vec{B}_d|/|\vec{B}_{IMF}|$, respectively, where $|\vec{B}_{IMF}|$ denotes the upstream IMF magnitude. The results are averaged into two-dimensional distributions with a spatial resolution of $0.2R_M \times 0.2R_M$ in MSE coordinates, and only bins containing more than 100 data points are retained.

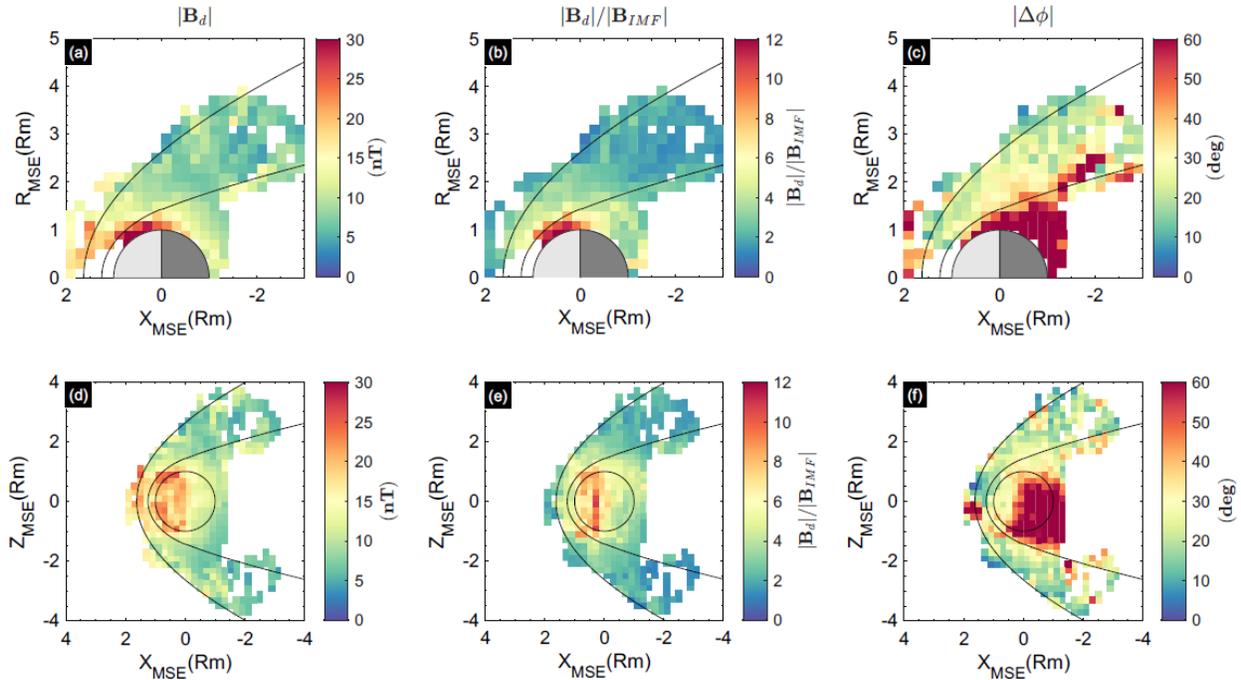

**Figure 3.** Spatial distributions of the downstream field strength $|\vec{B}_d|$, normalized field strength $|\vec{B}_d|/|\vec{B}_{IMF}|$, and clock angle departures (the absolute value of $\Delta\phi$, $|\Delta\phi|$). (a) - (c) show the distribution in the $X_{MSE} - R_{MSE}$ plane; (d) - (f) show the distribution in the $X_{MSE} - Z_{MSE}$ plane. The $X_{MSE} - Z_{MSE}$ plane has a $Y_{MSE}$ limit of $-1.5R_M \leq Y_{MSE} \leq 1.5R_M$.

Figures 3a and 3b show that both $|\vec{B}_d|$ and $|\vec{B}_d|/|\vec{B}_{IMF}|$ increase as approaching Mars on the dayside, forming a magnetic barrier (e.g., McComas et al., 1986; Zhang et al., 2022). This indicates strong magnetic field compression and mass-loading effects in this region (e.g., Tadlock et al., 2025). Within the nominal magnetosheath, bounded by the nominal bow shock (BS) and the magnetic pile-up boundary (MPB; Trotignon et al., 2006), both $|\vec{B}_d|$ and $|\vec{B}_d|/|\vec{B}_{IMF}|$ gradually decrease from the subsolar region toward the flanks.

Figure 3c illustrates the spatial distribution of $|\Delta\phi|$. In the outer magnetosheath near the nominal BS, $|\Delta\phi|$ remains within ~30°. Moving inward toward the MPB, $|\Delta\phi|$ increases to ~45°. Within the induced magnetosphere on the nightside magnetotail, $|\Delta\phi|$ typically exceeds 60°, consistent with the case study described earlier. Thus, the clock angle proxy method is generally applicable in the magnetosheath, most accurate near the bow shock, and not applicable within the induced magnetosphere. However, in the subsolar region, some $|\Delta\phi|$ unexpectedly exceed 45°. A re-examination of our database, along with two representative cases shown in Figures S1–S2 of the Supporting Information, suggests that these large $|\Delta\phi|$ values likely occur when the solar wind monitor enters the foreshock region, when numerous solar wind discontinuities are present, or when strong dayside crustal magnetic fields exert a remote influence. A detailed investigation of these possibilities is beyond the scope of this study. Nonetheless, these localized anomalies do not affect our results.

Figures 3d-3e reveals a sign of asymmetry in $|\vec{B}_d|$ and $|\vec{B}_d|/|\vec{B}_{IMF}|$, which is more evident in Figures S3a–S3b of the Supporting Information. Specifically, within the inner magnetosheath and magnetosphere, the values are slightly higher in the $+\vec{Z}_{MSE}$ hemisphere (which also corresponds to the +E hemisphere) than in the $-\vec{Z}_{MSE}$ (–E) hemisphere. This asymmetry is attributed to stronger mass loading associated with the plume (e.g., Zhang et al., 2022; Luhmann et al., 1985; Dubinin et al., 2019). Figures 3f and S3c show that $\Delta\phi$ also displays a pronounced E-hemisphere asymmetry. In the magnetosheath, $|\Delta\phi|$ is slightly larger in the +E hemisphere, while in the nightside magnetosphere, it becomes significantly larger in the –E hemisphere.

The E-hemisphere asymmetry of $|\Delta\phi|$ can be explained by the combined effects of "sinking" fields and tightly wrapped fields in the induced magnetosphere. In the terminator magnetosheath, magnetic field lines tend to sink toward the planetary shadow to maintain pressure balance (e.g., Perez-de-Tejada, 1986; Rong et al., 2014; Zhang et al., 2022). As illustrated in Figure S4, this sinking behavior leads to a vertical deflection of the field lines, imparting an upward or downward component that results in a nonzero Bz, thereby altering the clock angle $\phi$. Because the field strength is greater in the +E hemisphere (see Figure S3a), the associated magnetic pressure is larger, producing stronger sinking fields and thus a larger $|\Delta\phi|$ in this region. Furthermore, both observations and simulations show that, in the nightside –E hemisphere, field lines become tightly wrapped and irregular, deviating substantially from the idealized configuration of the induced magnetosphere (T. L. Zhang et al., 2010; Dubinin et al., 2019; Saunders & Russell, 1986). This naturally explains why $|\Delta\phi|$ is significantly larger in the nightside region of -E hemisphere. Overall, the clock-angle proxy method is generally applicable in the magnetosheath of both the +E and –E hemispheres.

Previous studies have shown that the induced magnetic field downstream of the shock exhibits distinct configurations on the quasi-parallel and quasi-perpendicular sides (e.g., Zhang et al., 2022; Tadlock et al., 2025; Shen et al., 2025). To assess how these differences influence the clock angle departures, we compute the local shock normal angle ($\theta_{Bn}$) for all downstream points using the bow shock model together with the upstream IMF. Specifically, we first determine the solar zenith angle (SZA) of each downstream point, then identify the corresponding location on the modeled bow shock surface at the same SZA. The shock normal vector at this location is then calculated, and $\theta_{Bn}$ is obtained as the angle between this shock normal vector and the upstream IMF. Figure 4a presents $|\Delta\phi|$ in the quasi-parallel side ($\theta_{Bn} < 45°$), while Figure 4b shows $|\Delta\phi|$ in the quasi-perpendicular side ($\theta_{Bn} > 45°$), and Figure 4c illustrates their difference. Overall, except for some localized regions, $|\Delta\phi|$ is consistently larger on the quasi-parallel side, with an average enhancement of approximately 15°. This is attributed to the more turbulent nature of the quasi-parallel side, where small-scale fluctuations and irregular magnetic field structures are more prevalent, resulting in larger clock angle departures.

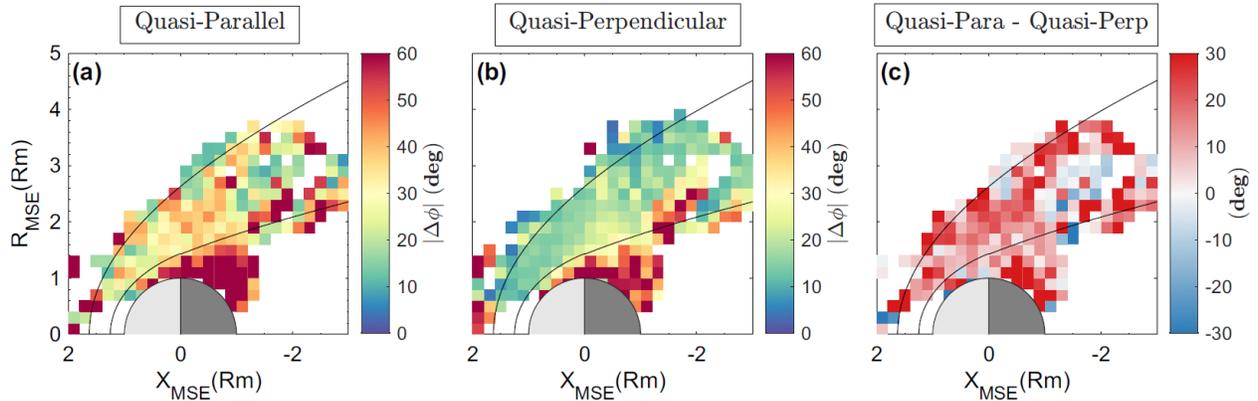

**Figure 4**. Spatial distributions of $|\Delta\phi|$ on the quasi-parallel side (a) and quasi-perpendicular side (b) of the bow shock. Panel (c) shows the difference in $|\Delta\phi|$ between the two regions.

### 3.2.1 Dependence on Solar Wind Dynamic Pressure

Previous studies have shown that the solar wind dynamic pressure (Pdy) can strongly influence the structure of the induced magnetosphere (e.g., Zhang et al., 2022; Y. Dong et al., 2019). Using our dataset, we quantitatively evaluate how Pdy affects $|\vec{B}_d|$, $|\vec{B}_d|/|\vec{B}_{IMF}|$ and $|\Delta\phi|$. To this end, we divide the dataset according to the median value of Pdy, which is 0.7 nPa. Data points with Pdy above the median are classified as the high-Pdy subset, while those below the median form the low-Pdy subset. The results are presented in Figure 5. The top panels (Figures 5a–5c) show $|\vec{B}_d|$, $|\vec{B}_d|/|\vec{B}_{IMF}|$ and $|\Delta\phi|$ for the low Pdy subset; the middle panels (Figures 5d–5f) display the corresponding quantities under high Pdy; and the bottom panels (Figures 5g–5i) illustrate the differences between the two subsets (high Pdy minus low Pdy).

From Figures 5a, 5d and 5g, we find that $|\vec{B}_d|$ increases significantly with increasing Pdy. However, the $|\vec{B}_d|/|\vec{B}_{IMF}|$ shows no clear dependence on Pdy (Figures 5b, 5e and 5h). This is because enhancements in Pdy are typically accompanied by corresponding increases in $|\vec{B}_{IMF}|$, as shown in Stergiopoulou et al. (2022). From Figures 5c, 5f and 5i, we find that $|\Delta\phi|$ near the terminator region close to Mars is slightly larger under low Pdy conditions. When Pdy is low, the induced magnetosphere expands outward, so more data points near the terminator fall within the induced magnetosphere. In contrast, under high Pdy, more data points in the same region lie within the magnetosheath. Therefore, the larger $|\Delta\phi|$ observed under low Pdy conditions can be explained by the greater fraction of measurements taken inside the induced magnetosphere.

Moreover, in the nightside region, we observe higher $|\Delta\phi|$ under high Pdy, which may suggest that draped field lines become more distorted, leading to enhanced $|\Delta\phi|$.

In a word, from a large-scale perspective, the difference in $|\Delta\phi|$ between low and high Pdy conditions is small, and $|\Delta\phi|$ generally remains below 40° in the magnetosheath, indicating that the clock angle proxy method is applicable in this region under both conditions. However, under high Pdy conditions, this method becomes unreliable in the magnetosheath region adjacent to the nominal MPB, where significant deviations are observed.

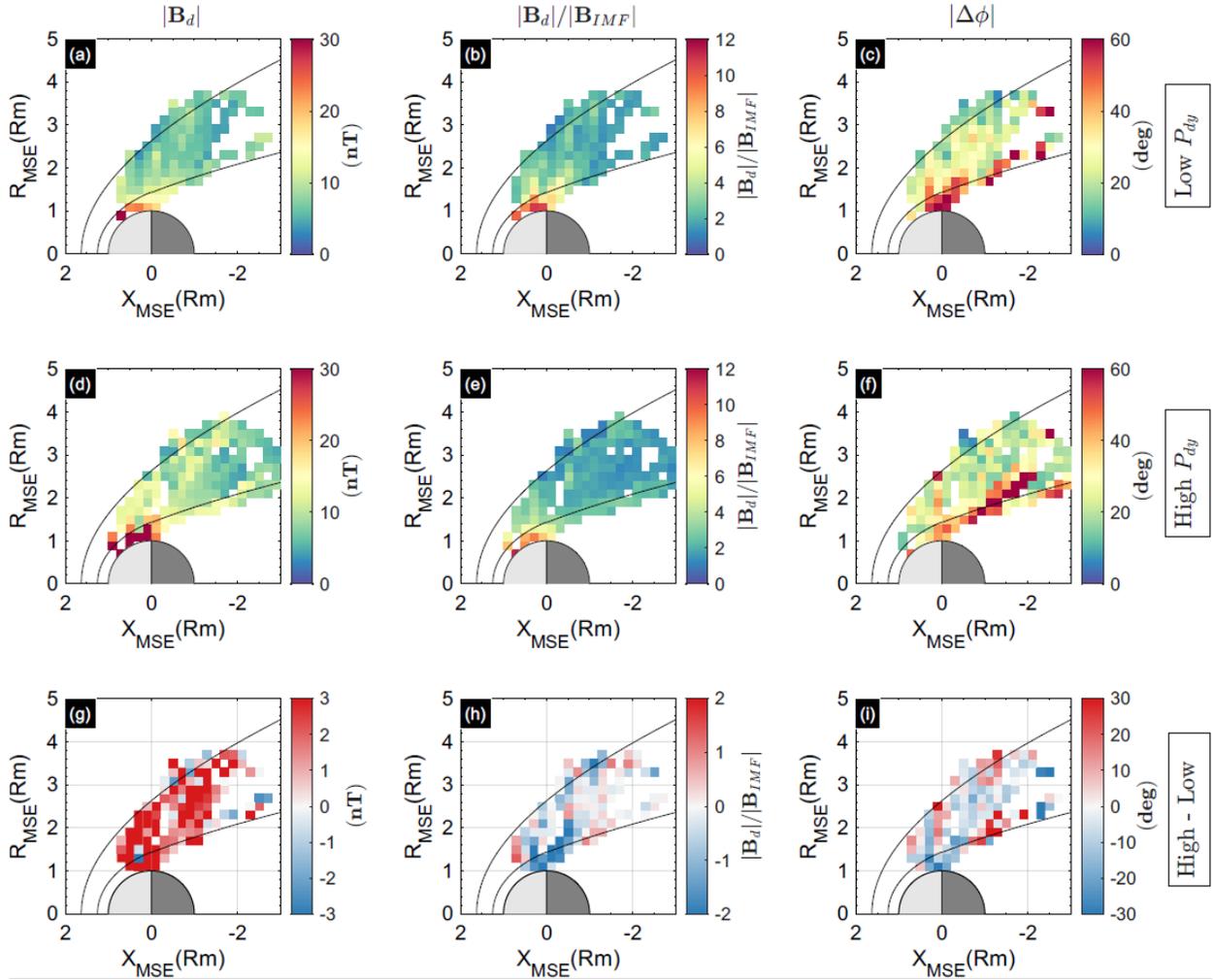

**Figure 5.** Spatial distributions of downstream field strength $|\vec{B}_d|$, normalized downstream field strength $|\vec{B}_d|/|\vec{B}_{IMF}|$, and clock angle departures $|\Delta\phi|$ under low (a-c) and high (d-f) solar wind dynamic pressures. The distributions are in the $X_{MSE} - R_{MSE}$ plane. (g–i) present the differences between the high Pdy and low Pdy cases.

### 3.2.2 Dependence on IMF Magnitudes

In this section, we investigate the dependence on the IMF magnitude ($|\vec{B}_{IMF}|$). The dataset is divided into high- and low-magnitude conditions using the median value of 3.02 nT. Figure 6 presents the spatial distributions of $|\vec{B}_d|$, $|\vec{B}_d|/|\vec{B}_{IMF}|$ and $|\Delta\phi|$ under these two conditions.

From Figures 6a, 6d and 6g, we see that larger $|\vec{B}_{IMF}|$ corresponds to larger $|\vec{B}_d|$. However, Figures 6b, 6e and 6h show that $|\vec{B}_d|/|\vec{B}_{IMF}|$ is lower under high $|\vec{B}_{IMF}|$ condition, indicating that the compressional effects are weaker when $|\vec{B}_{IMF}|$ is higher.

Figures 6c and 6f compare $|\Delta\phi|$ under low and high $|\vec{B}_{IMF}|$ conditions. In the magnetosheath, $|\Delta\phi|$ is lower under high $|\vec{B}_{IMF}|$ condition (see Figure 6i), particularly on the dayside and near the terminator. However, it is important to note that high $|\vec{B}_{IMF}|$ is typically accompanied by elevated Pdy (Stergiopoulou et al., 2022). As a result, the magnetosheath is compressed inward under high $|\vec{B}_{IMF}|$, whereas under low $|\vec{B}_{IMF}|$ condition it expands outward. Consequently, in Figure 6c, more data points within the nominal magnetosheath region fall inside the induced magnetosphere, where $|\Delta\phi|$ is larger. This behavior is consistent with the case discussed in Section 3.2.1. Therefore, we cannot conclude that $|\vec{B}_{IMF}|$ is negatively correlated with $|\Delta\phi|$, as the effect of Pdy is entangled with that of $|\vec{B}_{IMF}|$.

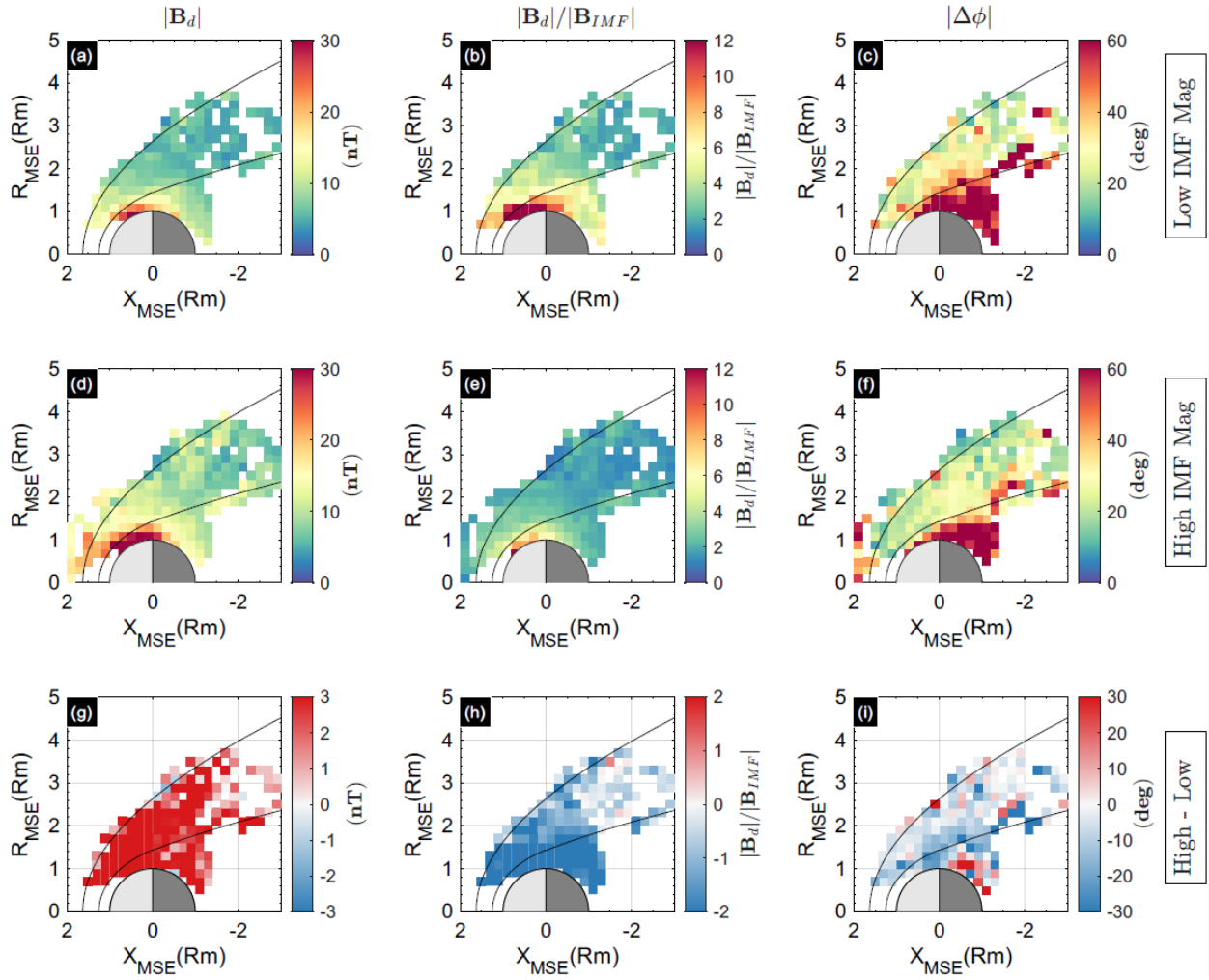

**Figure 6.** Spatial distributions of downstream field strength $|\vec{B}_d|$, normalized downstream field strength $|\vec{B}_d|/|\vec{B}_{IMF}|$, and clock angle departures $|\Delta\phi|$ under low (a-c) and high (d-f) IMF magnitude conditions. The distributions are in the $X_{MSE} - R_{MSE}$ plane. (g–i) present the differences between the high and low IMF magnitude conditions.

### 3.2.3 Dependence on IMF Cone Angle

In this section, we investigate the dependence on the IMF cone angle. We define the IMF cone angle, $\theta_{imf}$, as the angle between the IMF vector and the +X direction in the MSO/MSE coordinate system, under the assumption that the solar wind flows along +X. For simplicity, we consider only quasi-parallel and quasi-perpendicular IMF conditions. Quasi-parallel IMFs are defined as cases with $\theta_{imf} < 30°$ or $\theta_{imf} > 150°$, while quasi-perpendicular IMFs correspond to $60 < \theta_{imf} < 120°$.

Figure 7 compares $|\vec{B}_d|$, $|\vec{B}_d|/|\vec{B}_{IMF}|$ and $|\Delta\phi|$ under quasi-perpendicular and quasi-parallel IMF conditions. Figures 7a and 7d, together with Figure 7g, show that $|\vec{B}_d|$ are clearly stronger under quasi-perpendicular IMF conditions compared with quasi-parallel IMF conditions. However, Figures 7b, 7e, and 7h reveal that the $|\vec{B}_d|/|\vec{B}_{IMF}|$ does not show a strong dependence on the IMF cone angle. To understand this behavior, we re-examined the dataset and found that quasi-parallel IMF conditions are often associated with lower $|\vec{B}_{IMF}|$ (see Figure S5). This implies that the effects of IMF strength and IMF cone angle are intertwined. As demonstrated in the previous subsection, weaker $|\vec{B}_{IMF}|$ tends to increase the $|\vec{B}_d|/|\vec{B}_{IMF}|$. Therefore, the lack of a clear correlation between $|\vec{B}_d|/|\vec{B}_{IMF}|$ and IMF cone angle suggests that the quasi-parallel IMF tends to suppress $|\vec{B}_d|/|\vec{B}_{IMF}|$. This is consistent with the expectations: quasi-perpendicular IMF enables more efficient mass loading and magnetic field amplification, leading to a larger compression ratio downstream (e.g., Q. Zhang et al., 2023; Chen et al., 2025).

Figures 7c, 7f and 7i show that $|\Delta\phi|$ is much larger in the magnetosheath under quasi-parallel IMF conditions than under quasi-perpendicular IMF conditions. Under quasi-parallel IMF, even within the magnetosheath, half of the bins exhibit $|\Delta\phi|$ values exceeding 50° (see Figure 7c), indicating that the clock angle proxy method is invalid in this case. This is because, under quasi-parallel IMF, the magnetosheath evolves into a quasi-parallel state, characterized by strong turbulence and irregular magnetic field structures that deviate from the classical draping pattern (e.g., Shen et al., 2025; Tadlock et al., 2025). Although weaker IMF strength can also increase $|\Delta\phi|$ as shown in the previous subsection, the enhancement seen in Figure 7 is much larger than that in Figure 6, indicating that the quasi-parallel configuration itself also contributes to the increased $|\Delta\phi|$.

Within the induced magnetosphere, we find that $|\Delta\phi|$ is generally larger than 50°, regardless of the IMF cone angle. This implies that the clock angle proxy method is only reliable in the magnetosheath under quasi-perpendicular IMF conditions.

manuscript submitted to *JGR: Space Physics*
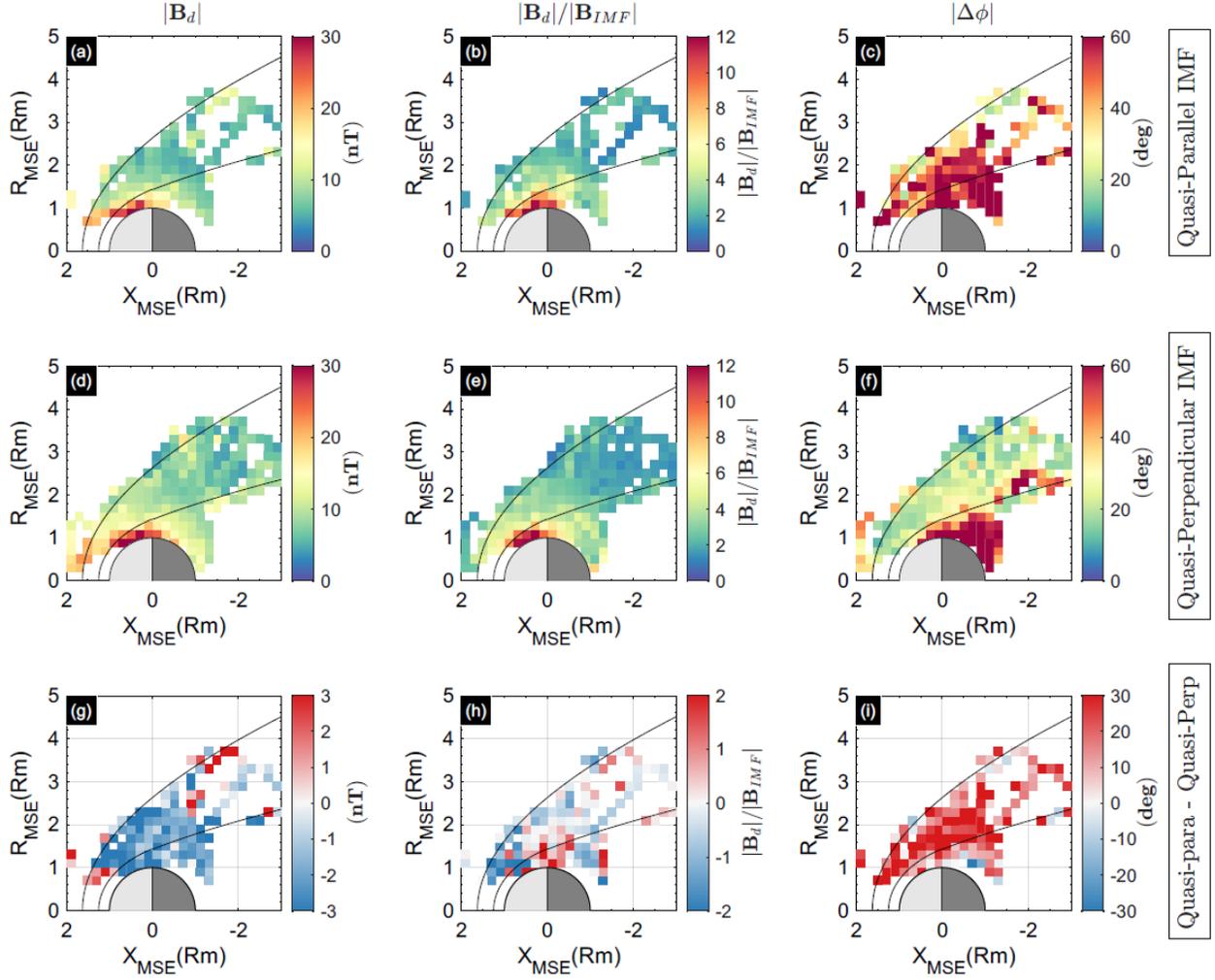

**Figure 7.** Spatial distributions of downstream field strength $|\vec{B}_d|$, normalized downstream field strength $|\vec{B}_d|/|\vec{B}_{IMF}|$, and clock angle departures $|\Delta\phi|$ under quasi-parallel (a-c) and quasi-perpendicular (d-f) IMF conditions. The distributions are in the $X_{MSE} - R_{MSE}$ plane. (g–i) present the differences between the quasi-parallel and quasi-perpendicular cases.

### 3.2.4 Dependence on $P_{dy}/P_B$ and Mach Number

From the above analysis, we find that both Pdy and $|\vec{B}_{IMF}|$ exert similar influences on $|\Delta\phi|$, likely because increases in Pdy are often accompanied by enhancements in $|\vec{B}_{IMF}|$. To further disentangle their relative roles, we examine the behavior of $|\vec{B}_d|$, $|\vec{B}_d|/|\vec{B}_{IMF}|$ and $|\Delta\phi|$ under

different $P_{dy}/P_B$ conditions, where $P_B$ denotes the magnetic pressure of the IMF, defined as $B_{IMF}^2/2\mu_0$, where $\mu_0$ is the vacuum permeability. Specifically, we divide the dataset into two regimes based on the median value of $P_{dy}/P_B$: high $P_{dy}/P_B$ (greater than 152.2) and low $P_{dy}/P_B$ (less than 152.2) conditions.

Figures 8a, 8d and 8g reveal that $|\vec{B}_d|$ is slightly higher under low $P_{dy}/P_B$ condition, whereas Figures 8b, 8e, and 8h indicate that $|\vec{B}_d|/|\vec{B}_{IMF}|$ is greater under high $P_{dy}/P_B$ condition. This suggests that $|\vec{B}_{IMF}|$ primarily determines the absolute strength of the draped magnetic field, while dynamic pressure contributes to enhancing the compression of the field lines. In contrast, stronger $P_B$ or $|\vec{B}_{IMF}|$ tends to reduce the relative compression.

Furthermore, Figures 8c, 8f, and 8i show that $|\Delta\phi|$ is significantly larger under the high $P_{dy}/P_B$ condition. This implies that elevated Pdy tends to increase the irregularity or distortion of the draped magnetic field lines, while stronger upstream magnetic fields act to suppress such deviations. It is worth noting that $P_{dy}/P_B = M_A^2$, where $M_A$ is the Alfvénic Mach number. Therefore, higher $M_A$ values correspond to greater deviations in the draped magnetic field, consistent with our observations. Therefore, we conclude that the clock angle proxy method is more reliable under low $M_A$ or $P_{dy}/P_B$ conditions, where the draped field lines remain more organized and less distorted.

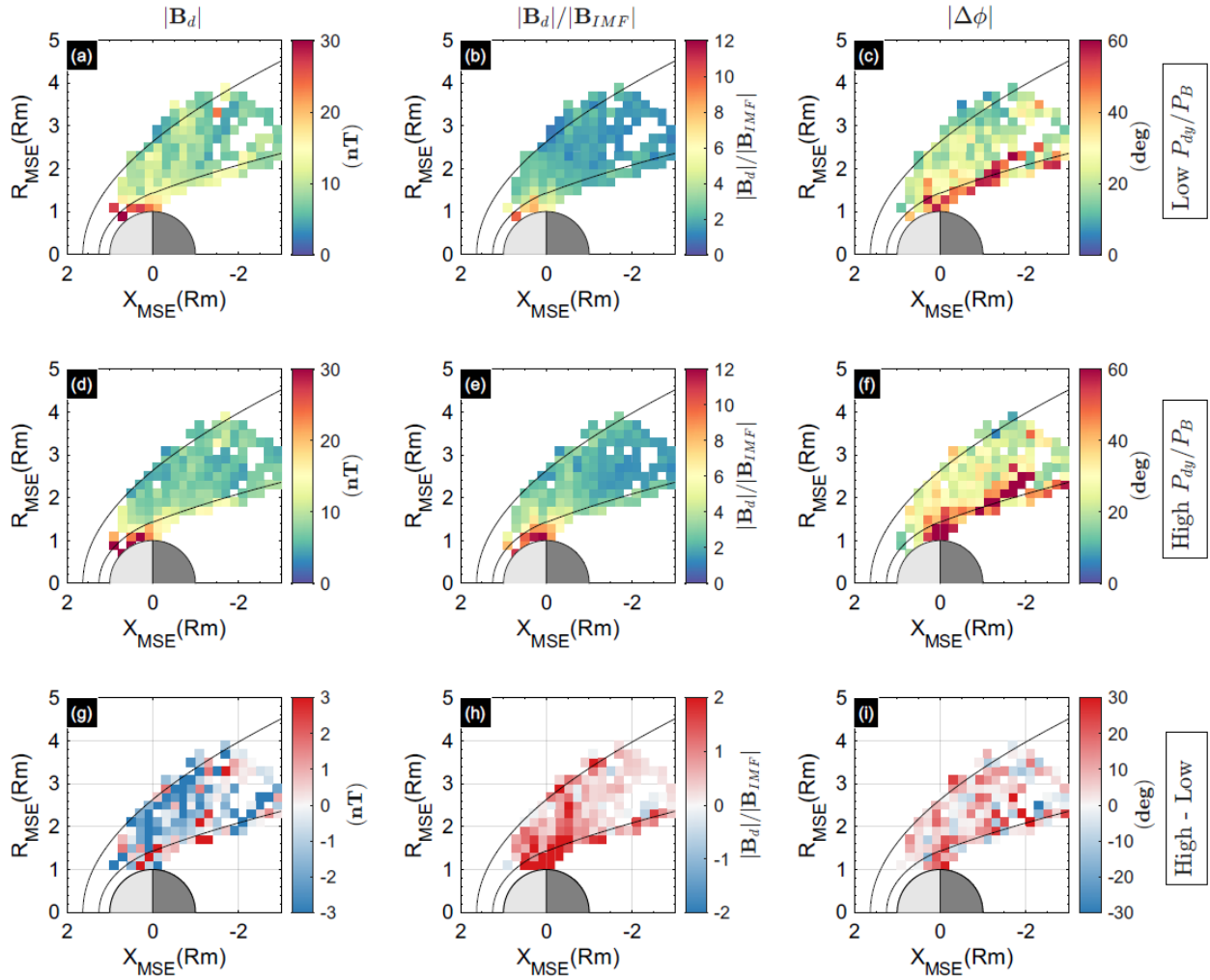

**Figure 8.** Spatial distributions of downstream field strength $|\vec{B}_d|$, normalized downstream field strength $|\vec{B}_d|/|\vec{B}_{IMF}|$, and clock angle departures $|\Delta\phi|$ under low (a-c) and high (d-f) $P_{dy}/P_B$ conditions. (g–i) present the differences between the low and high $P_{dy}/P_B$ conditions.

## 4 Conclusions and Discussions

In this study, we take advantage of the simultaneous observations from MAVEN and Tianwen-1 to investigate Mars' induced magnetic field and the associated clock angle departures, with the addition of real-time upstream solar wind conditions. In previous analyses, the characterization of the induced magnetic fields was based on the assumption that solar wind conditions remained steady throughout the orbital period of a single spacecraft. However, this assumption introduces uncertainty, particularly when the spacecraft is located in the downstream region, since the actual upstream solar wind may have changed significantly during that time. The dual-spacecraft configuration now allows us to directly incorporate contemporaneous upstream measurements and the downstream induced magnetic fields near Mars. This configuration effectively eliminates the uncertainty arising from time-variable upstream drivers and enables a more accurate assessment of the properties of the induced magnetic fields and their variations. The key findings of this work are summarized as follows:

(1) Both enhanced Pdy and $|\vec{B}_{IMF}|$ increase the downstream magnetic field strength ($|\vec{B}_d|$); they have opposite effects on the compression ratio ($|\vec{B}_d|/|\vec{B}_{IMF}|$). Specifically, higher Pdy leads to a stronger compression ratio, whereas stronger $|\vec{B}_{IMF}|$ or $P_B$ tend to reduce it. Taken together, this means that larger $M_A$ correspond to stronger compression ratios. Furthermore, we find that the compression ratio ($|\vec{B}_d|/|\vec{B}_{IMF}|$) is lower under quasi-parallel IMF conditions than under quasi-perpendicular conditions, indicating that mass loading and the magnetic field amplification are more effective when the IMF is oriented more perpendicularly to the solar wind flow.

(2) The clock angle departures ($|\Delta\phi|$) are generally smaller than 45° in the magnetosheath and increase gradually from the bow shock toward the induced magnetosphere. Within the induced magnetosphere, $|\Delta\phi|$ typically exceeds 60°, indicating that the clock angle proxy method is not valid in this region. In addition, we find that $|\Delta\phi|$ exhibits a clear E-asymmetry and a four-quadrant structure, resulting from the sinking fields and tightly wrapped field lines.

(3) We find that enhanced Pdy increase $|\Delta\phi|$, whereas stronger $|\vec{B}_{IMF}|$ reduces it. This suggests that elevated dynamic pressure tends to distort the draped magnetic field lines, while stronger upstream magnetic fields act to stabilize them. Consequently, large $M_A$ are associated with stronger departures in clock angle. Moreover, $|\Delta\phi|$ is generally larger in the quasi-parallel magnetosheath than in the quasi-perpendicular region, which is due to the more turbulent and irregular magnetic structures in the quasi-parallel regime. Under quasi-parallel IMF conditions, $|\Delta\phi|$ becomes particularly large. Therefore, we conclude that the clock angle proxy method is invalid under quasi-parallel IMF conditions, whereas under quasi-perpendicular IMF conditions it is more reliable, especially when Pdy is low, $|\vec{B}_{IMF}|$ is high, and $M_A$ is small.

The above results are derived from simultaneous upstream and downstream observations and therefore neglect the propagation delay of the upstream IMF to the downstream region. Under steady solar wind conditions, this delay is negligible. In contrast, when an IMF discontinuity alters the upstream solar wind, the downstream magnetic field typically responds after a delay of 8 seconds to 11 minutes, depending on the observation location and the structure of the discontinuity (e.g., Romanelli et al., 2019; Guo et al., 2025). During such intervals, the reliability of the derived results may be compromised. Nevertheless, these disturbed intervals constitute only a small portion of our large dataset, and their influence is largely mitigated through statistical averaging.

Figures 3 and S3 reveal an E-hemisphere asymmetry in both magnetic field strength and clock angle departures, consistent with previous studies that identified the presence of sinking and wrapped magnetic fields (e.g., Zhang et al., 2022). These findings suggest that such field configurations are intrinsic features of Mars' induced magnetosphere. However, it remains unclear whether these structures vary in response to changing solar wind conditions. Our ability to investigate this relationship is limited by the spatial coverage of the current 10 months dataset, which does not adequately sample the deep induced magnetotail region (see Figures 3–8). Additionally, previous studies have shown that the induced magnetic fields at Venus vary with solar EUV levels, exhibiting distinct characteristics during solar minimum and maximum (Xiao et al., 2018). However, our dataset is also insufficient to resolve the long-term effects of solar EUV variability on the induced magnetic fields and clock angle departures. Therefore, it is essential to extend the dataset to enable a more comprehensive investigation of the Martian induced magnetic fields and their responses to varying solar wind and EUV conditions, under real-time upstream monitoring.

**Data and Code Availability Statement**

All data used in this paper are public. The MAVEN MAG data is publicly archived at https://pds-ppi.igpp.ucla.edu/mission/MAVEN/Magnetometer (Connerney, 2024). The MAVEN SWIA data is publicly archived at https://pds-ppi.igpp.ucla.edu/mission/MAVEN/maven/SWIA (Halekas, 2024). The Tianwen-1 MOMAG data sets are publicly available at https://moon.bao.ac.cn/web/zhmanager/mars1 and http://space.ustc.edu.cn/dreams/tw1_momag/. The dataset used in this paper is available on Zenodo (Zhang & Cheng, 2025): https://doi.org/10.5281/zenodo.17624009.

## Acknowledgments

This work was partially supported by NASA grant NNH10CC04C through the MAVEN Project, NASA grants 80NSSC23K0911 and 80NSSC24K1843, and the Alfred P. Sloan Research Fellowship.

## Conflict of Interest Statement

The authors have no conflicts of interest to disclose.